\documentclass[twocolumn]{aastex631}

\graphicspath{{./}{figures/}}
\usepackage{amsmath}

\usepackage{CJK}

\newcommand\sref[1]{\hyperref[#1]{\S~\ref*{#1}}}
\newcommand\fref[1]{\hyperref[#1]{Fig.~\ref*{#1}}}
\newcommand\Eqref[1]{Eq.~(\hyperref[#1]{\ref*{#1}})}
\newcommand\eeqref[1]{Eq.~\hyperref[#1]{\ref*{#1}}}
\newcommand\tref[1]{\hyperref[#1]{Table~\ref*{#1}}}
\newcommand\aref[1]{\hyperref[#1]{Appendix~\ref*{#1}}}

\shorttitle{Bridging Scales: Galactic nucleus to cosmic environment}
\shortauthors{Su et al.}

\begin{document}
\begin{CJK}{UTF8}{mj}
\title{Bridging Scales: Multi-Zone GRMHD-Informed Anisotropic AGN Feedback in Galaxies}
\author[0000-0003-1598-0083]{Kung-Yi Su}
\affiliation{Department of Physics \& Astronomy and Center for Interdisciplinary Exploration and Research in Astrophysics(CIERA), Northwestern University, 1800 Sherman Ave, Evanston, IL 60201, USA}
\email{kungyisu@gmail.com}

\author[0000-0002-2858-9481]{Hyerin Cho (조혜린)}
\affiliation{TAPIR, Mailcode 350-17, California Institute of Technology, Pasadena, CA 91125, USA}
\affiliation{Center for Astrophysics $\vert$ Harvard \& Smithsonian, 60 Garden Street, Cambridge, MA 02138, USA}
\affiliation{Black Hole Initiative at Harvard University, 20 Garden Street, Cambridge, MA 02138, USA}

\author[0000-0002-5554-8896]{Priyamvada Natarajan}
\affiliation{Black Hole Initiative at Harvard University, 20 Garden Street, Cambridge, MA 02138, USA}
\affiliation{Department of Astronomy, Yale University, Kline Tower, 266 Whitney Avenue, New Haven, CT 06511, USA}
\affiliation{Department of Physics, Yale University, P.O. Box 208121, New Haven, CT 06520, USA}

\author[0000-0002-4900-6628]{Claude-Andr\'e Faucher-Gigu\`ere }
\affiliation{Department of Physics \& Astronomy and Center for Interdisciplinary Exploration and Research in Astrophysics(CIERA), Northwestern University, 1800 Sherman Ave, Evanston, IL 60201, USA}

\author[0000-0002-1919-2730]{Ramesh Narayan}
\affiliation{Center for Astrophysics $\vert$ Harvard \& Smithsonian, 60 Garden Street, Cambridge, MA 02138, USA}
\affiliation{Black Hole Initiative at Harvard University, 20 Garden Street, Cambridge, MA 02138, USA}

\author[0000-0001-5287-0452]{Angelo Ricarte}
\affiliation{Black Hole Initiative at Harvard University, 20 Garden Street, Cambridge, MA 02138, USA}
\affiliation{Center for Astrophysics $\vert$ Harvard \& Smithsonian, 60 Garden Street, Cambridge, MA 02138, USA}

\author[0000-0002-1996-0445]{Antonio J. Porras-Valverde}
\affiliation{Department of Astronomy, Yale University, Kline Tower, 266 Whitney Avenue, New Haven, CT 06511, USA}

 \author[0000-0002-0393-7734]{Cora Prather} 
\affiliation{Black Hole Initiative at Harvard University, 20 Garden Street, Cambridge, MA 02138, USA}

\author[0000-0003-3729-1684]{Philip F. Hopkins}
\affiliation{TAPIR, Mailcode 350-17, California Institute of Technology, Pasadena, CA 91125, USA}

\begin{abstract}
We present a galaxy-scale black hole (BH) accretion and feedback model incorporating results from multi-zone general relativistic magnetohydrodynamic (GRMHD) simulations spanning the event horizon to the Bondi radius. The model replaces a fixed suppression of the Bondi rate with a nuclear-environment-dependent factor, $\kappa^{\rm hot}_{\rm cho}$, set by the local ratio of Bondi to gravitational radius, and distributes outflow mass, kinetic energy, and thermal energy according to the angular structure measured in the large-dynamic-range GRMHD calculations. We apply it to idealized isolated simulations of an M87*-like massive elliptical and a Milky Way/Sgr A*-like galaxy, spanning over three orders of magnitude in BH mass without tuning accretion or feedback efficiencies to match observables. In the M87*-like system, the GRMHD-informed angle-dependent feedback model suppresses late-time stellar mass growth more effectively than the isotropic reference model, despite similar long-term BH accretion rates. Varying the uncertain mass loading of the collimated polar component has a modest impact on BH growth and star formation, indicating that feedback anisotropy matters more than its detailed mass loading. We find that $\kappa^{\rm hot}_{\rm cho}$ varies with nuclear environment and time, yielding BH growth histories markedly different from fixed-suppression models. For Sgr A*, this variable factor yields an accretion rate more consistent with Event Horizon Telescope (EHT) constraints than a fixed, suppressed Bondi model. These results show that nuclear-environment-dependent accretion and feedback angular structure affect BH-galaxy evolution, requiring subgrid models to include more than a universal Bondi suppression factor and a scalar feedback efficiency.

\end{abstract}


\section{Introduction}
\label{S:intro}

Understanding how supermassive black holes (SMBHs) are fueled, how much of the gas supplied to their sphere of influence ultimately reaches the event horizon, and how the energy released by accretion couples back to the host galaxy is an intrinsically multiscale problem. Gas is supplied on circumgalactic and galactic scales, transported through the nuclear region, and ultimately accreted on scales of the order of the gravitational radius, while the associated feedback can propagate back outward to affect the interstellar medium (ISM), circumgalactic medium (CGM), and, in sufficiently powerful systems, scales comparable to the virial radius. The feeding and feedback problem is therefore fundamentally bi-directional: the large-scale environment determines the boundary conditions of the accretion flow, while processes operating close to the black hole (BH) determine how much of that inflow is accreted and how much mass, momentum, and energy are returned to the host.

The enormous dynamic range involved has traditionally required a separation between simulations of BH accretion and simulations of galaxy evolution. General relativistic magnetohydrodynamic (GRMHD) calculations can resolve the dynamics of magnetized plasma close to the event horizon and power BH feedback self-consistently, but they cannot ordinarily follow the surrounding galactic atmosphere over their relevant spatial and temporal scales \citep[e.g.,][]{Gammie2003,Tchekhovskoy2011,Porth:2019,Chatterjee:2023}. Conversely, galaxy-scale and cosmological simulations can follow gas dynamics from the ISM to the CGM and beyond, but cannot resolve the gravitational-radius-scale accretion flow. They therefore require sub-resolution prescriptions to connect resolved gas properties to the BH accretion rate and to specify the energetics, mass loading, and geometry of the resulting feedback, with model parameters -- often implicitly or explicitly dependent on BH mass -- tuned to match observed scaling relations \citep[e.g.,][]{Li2014,Fiacconi2018,AnglesAlcazar+Zoom_2021,Talbot2021,Weinberger+2023}. Prescriptions of this kind are used in major simulation suites \citep[e.g.,][]{Sijacki2015,Rosas-Guevara2016,AngleAlcazar+2017,Weinberger2018,Ricarte2019,Ni2022,Wellons2023,2024ApJ...973..149B,2024MNRAS.531.4311B}. While some simulations \citep[e.g.,][]{Hopkins+ForgeInFIRE1+2024, Guo+2024} have adopted zoom-in methods to resolve the flow down to BH-relevant scales, the outward propagation of the BH feedback to galactic scales remains inaccessible. A central challenge is consequently to replace phenomenological assumptions in these prescriptions, wherever possible, with physically motivated relations derived from calculations that bi-directionally connect the relevant scales directly. 

We have been pursuing this ``bridging scales'' problem through a sequence of simulations designed to connect horizon-scale accretion physics to the galactic environment. In the first study of this series, \citet{Cho+2023} developed a novel multi-zone GRMHD method capable of following magnetized accretion over approximately seven orders of magnitude in radius. Applied to an idealized Bondi flow problem with a Bondi radius $R_{\rm B}\sim2\times10^{5}\,r_g$ for a non-spinning BH, these initial calculations showed that magnetic fields qualitatively alter the classical Bondi solution. The inner flow becomes strongly magnetized, the density profile approaches $\rho\propto r^{-1}$ rather than the Bondi scaling of $\rho\propto r^{-3/2}$, and the accretion rate reaching the BH horizon is suppressed by more than two orders of magnitude relative to the nominal Bondi rate. Even in the absence of relativistic jets, these simulations also revealed continuous outward energy transport at an efficiency $\eta=0.02$, driven by magnetic reconnection and turbulent motions near the BH, explicitly demonstrating that magnetic processes arising close to the horizon can modify the flow far beyond the relativistic regime. 

Building on this framework, \citet{Cho+2024} extended the multi-zone approach to obtain dynamical steady states over as many as eight decades in radius and connected the inner GRMHD calculation to initial conditions and an external potential drawn from a galaxy-scale model identical to the M87* case studied in this work. For a non-spinning BH, these calculations again found an extended, strongly magnetized accretion flow and an accretion rate far below the analytic Bondi value. More importantly for applications on unresolved scales, simulations spanning a wide range of Bondi radii, $R_{\rm B}\sim10^{2}$--$10^{7}\,r_g$, yielded the approximate relation
\begin{equation}
    \frac{\dot{M}_{\rm BH}}{\dot{M}_{\rm Bondi}}
    \simeq
    \left(\frac{R_{\rm B}}{6\,r_g}\right)^{-1/2}.
\end{equation}
This result provides a physically motivated mapping between the resolved thermodynamic state at the Bondi scale and the mass flux reaching the BH. Because $R_{\rm B}/r_g$ depends on the sound speed and relative velocity of the surrounding gas, the effective suppression of Bondi accretion is not a universal numerical factor but instead responds to the nuclear environment in which the black hole is embedded.

Following the multizone method, similar bridging scale techniques have since been implemented in Cartesian \citep[cyclic zoom;][]{Guo+2025} and Lagrangian \citep[time dilation;][]{Hopkins+2025timedialation} codes, using  techniques adapted for each. In particular, \citet{Guo+2025} successfully reproduced the non-spinning BH results of \citet{Cho+2023,Cho+2024}, offering the first cross-validation of large-dynamic-range simulations that had previously been out of reach.

The next step was to determine how these horizon-to-Bondi-scale results affect the evolution of a galactic atmosphere. \citet{Su+2025Bridging} embedded the GRMHD-derived accretion suppression and derived feedback efficiency into high-resolution, non-cosmological MHD simulations of M87$^\ast$- and Sgr A$^\ast$-like systems, including the full galaxies and halo gas, using the FIRE-2 framework. These calculations showed that the suppressed-Bondi prescription substantially reduces BH growth. 
They also revealed an important tension between the feedback efficiency inferred for a non-spinning BH, $\eta=0.02$ relative to $\dot{M}_{\rm BH}c^2$, which appeared generally insufficient to regulate the cooling flows strongly enough to simultaneously reproduce the inferred black hole accretion rates and star-formation constraints in M87$^\ast$ and Sgr A$^\ast$. Feedback efficiencies of order $\gtrsim15\%$ were required in those isolated galaxy simulations, suggesting that an additional source of feedback energy associated with BH spin could be dynamically important and needs to be included in the computational framework.

This possibility was subsequently explored by \citet{Cho+2025}, who generalized the multi-zone GRMHD calculations to a rapidly spinning BH ($a_*=0.9$), and later in \citet{Cho2026}, who presented a comprehensive library of simulations spanning a wide range of BH spins and Bondi radii, providing spin-dependent subgrid prescriptions from first principles. In these calculations, rapidly rotating BHs launch relativistic, highly magnetized polar outflows and attain substantially larger feedback efficiencies than their non-spinning counterparts. At the same time, the accretion rate continues to exhibit an approximate $\dot{M}_{\rm BH}/\dot{M}_{\rm Bondi}\propto R_{\rm B}^{-1/2}$ dependence over the Bondi-radius range explored. These results separate two physically distinct aspects of the multiscale problem: the fraction of the Bondi supply that reaches the BH is set primarily by the thermodynamic and kinematic state of the surrounding gas through $R_{\rm B}/r_g$, whereas the efficiency and structure of the feedback are strongly influenced by the state of the inner accretion flow and, in particular, by BH spin.

More recently, \citet{Su+2026_zoom_in} carried this GRMHD-informed suppressed-Bondi framework into cosmological magnetohydrodynamic zoom-in simulations of halos with masses of order $10^{14}\,M_\odot$. These calculations demonstrated that the consequences of suppressed accretion extend beyond idealized isolated systems. In the suppressed-Bondi regime, BHs with masses below $\sim10^{7}\,M_\odot$ grow inefficiently via gas accretion, while at higher masses, their evolution depends sensitively on feedback efficiency. Low-spin systems can continue to grow without effectively quenching their hosts, whereas high-spin systems can quench star formation efficiently but subsequently become starved of fuel. The cosmological calculations, therefore, expose a broader tension between producing sufficient SMBH growth and sufficiently strong maintenance-mode feedback within a single hot-mode accretion prescription. This tension is reflected in the significant disagreement amongst cosmological simulations in reproducing the low-mass slope of the quenched central stellar mass function, a key observational diagnostic of AGN feedback and BH-galaxy coevolution \citep{Porras-ValverdeForbes2025}.

A parallel line of work using semi-analytical models has also incorporated horizon-to-Bondi-scale GRMHD simulation results.
\citet{Porras-Valverde2026Bridgingscales} incorporated the accretion suppression prescription into the {\sc Dark Sage} semi-analytic model, finding that restricting the suppression to sub-Eddington accretion ($f_{\rm Edd}<3\times10^{-3}$) and adopting distinct feedback efficiencies at low and high Eddington rates reproduces the observed local distribution of galaxy stellar masses and black hole masses while preserving the rapid super-Eddington growth inferred at $z>3$.
Additionally, Porras-Valverde
et al. (2026c, in prep.) implemented spin-dependent feedback under magnetically arrested disk conditions, finding that SMBH in massive galaxies spin down to $a_\star \sim 0.035$, effectively suppressing both radiative efficiency and jet power. An additional source of feedback \citep[e.g.,][]{Cho+2023} is required to reproduce the high-mass end of the observed stellar mass function.

Similarly, \citet{Su+2026_zoom_in} also emphasize the likely need for a multi-regime description, including hot vs. cold inflows and sub- vs. super-Eddington accretion, as well as the effects of anisotropic and spin-dependent feedback and galaxy mergers, all of which contribute at different stages of SMBH assembly. Indeed, semianalytic modeling of SMBH mass assembly across $M_{\rm BH}\sim10^{6}$--$10^{10}\,M_\odot$ finds that secular disk instabilities and merger-driven cold accretion dominate the growth budget for most black holes, with hot-mode accretion playing a secondary role \citep{PorrasValverde2026MassAssembly}.

Taken together, this sequence of calculations has established a physically motivated connection between the resolved gas properties at the Bondi scale, the accretion rate reaching the BH, and the total energetic efficiency of the resulting feedback. An important aspect of the nuclear-scale solution, however, has so far received less attention in galaxy-scale applications, which is that the feedback is not characterized by a single scalar efficiency. The outflow emerging from the GRMHD calculation is highly anisotropic. Its mass flux, kinetic-energy flux, and thermal-energy flux vary strongly with polar angle, with the highest-specific-energy material concentrated toward the rotation axis \citep{Cho+2025,Cho2026}. Thus, even if the total feedback power were known exactly, it would not uniquely determine the response of the host galaxy. The subsequent evolution depends on where the energy and momentum are deposited, which in turn depends on the angular distribution, velocity, mass loading, and thermal content of the outflow.

This geometrical information may be particularly important on larger galactic scales. A fast, low-density polar outflow can propagate to large radii before dissipating a substantial fraction of its energy, whereas material launched with a broader angular distribution can interact more directly with the central, denser gas responsible for feeding a cooling flow \cite[e.g.,][]{Su+2021,Su+JetHaloMass2024}. The same integrated feedback power can therefore produce very different large-scale effects depending on its angular structure. Conversely, details of the most highly collimated polar component may become dynamically unimportant if the subsequent interaction with the surrounding atmosphere erases the differences before they reach galactic scales. Determining which aspects of the nuclear outflow survive this propagation requires following the feedback self-consistently into a resolved galactic environment.

The present work addresses the next step in the bridging-scales program. We take both the nuclear-environment-dependent accretion prescription and the angular structure of the outflow, inferred from the multi-zone GRMHD calculations, and embed them in galaxy-scale simulations of M87$^\ast$- and Sgr A$^\ast$-like systems. The black-hole accretion rate is written as
\begin{equation}
    \dot{M}_{\rm BH}
    =
    \kappa^{\rm hot}_{\rm cho}\dot{M}_{\rm Bondi},
\end{equation}
where $\kappa^{\rm hot}_{\rm cho}$ is determined by the local Bondi radius in units of the gravitational radius, and the mass, kinetic-energy, and thermal-energy fluxes of the corresponding feedback are distributed according to the angular structure measured in the nuclear-scale calculations. Our aim is therefore not to introduce another phenomenological jet prescription or artificially assume isotropic feedback, but to directly propagate the self-consistent horizon-to-Bondi-scale solution outward and determine which components of that solution remain dynamically important once the outflow interacts with the multiphase galactic environment.

We use these simulations to address three closely related questions. First, how does allowing $\kappa^{\rm hot}_{\rm cho}$ to respond self-consistently to the thermodynamic and kinematic state of the resolved nuclear gas alter black-hole growth relative to models adopting a fixed suppression of the Bondi rate? Second, at fixed underlying accretion and feedback energetics, how strongly does the response of the host depend on the GRMHD-informed angular feedback structure of the outflow rather than simply on its integrated power? And finally, how sensitive are the galaxy-scale results to the uncertain mass loading and specific energy of the most highly collimated polar component?

To address these questions, we have performed idealized isolated-galaxy simulations representative of two very different environments: an M87$^\ast$-like massive elliptical containing a $6.5\times10^{9}\,M_\odot$ BH and a Milky Way analog hosting a $4.3\times10^{6}\,M_\odot$ Sgr A$^\ast$-like BH. The two systems differ by more than three orders of magnitude in BH mass and also differ substantially in halo mass, virial temperature, gas content, and angular momentum. The Sgr A$^\ast$ analog additionally contains rotationally supported stellar and gaseous disks and is explored with different CGM normalizations. We emphasize, however, that we do not independently tune the underlying BH accretion and feedback model for each host to obtain the desired galaxy or black hole properties. Instead, different effective values of $\kappa^{\rm hot}_{\rm cho}$ result in the emergence of different  feedback responses even when applying the same GRMHD-informed prescription to their varied resolved environments.

Therefore, we extend the bridging-scales program in both directions with the present calculations. The multi-zone GRMHD simulations follow the influence of large-scale boundary conditions inward from the Bondi scale to the horizon and determine the associated mass and energy fluxes flowing back outward. Previous galaxy-scale and cosmological calculations \citep{Su+2025Bridging,Su+2026_zoom_in} have tested the consequences of the integrated accretion suppression and feedback efficiency inferred from these simulations. Here, we additionally carry the \emph{angular structure} of the nuclear outflow into the host galaxy and allow its interaction with the surrounding atmosphere to determine the resulting coupling. The broader goal is to move from subgrid models described primarily by scalar efficiencies toward a more direct exchange of physically resolved information between simulations operating on otherwise inaccessible spatial and temporal scales.

The remainder of this paper is organized as follows. In \sref{S:methods}, we describe the galaxy models, the GRMHD-informed BH accretion prescription, and the implementation of the angle-dependent outflow. We then compare the resulting BH growth, star formation, and gas evolution for isotropic and GRMHD-informed angular feedback models in the M87$^\ast$ and Sgr A$^\ast$ analogs, and examine the consequences of replacing the nuclear-environment-dependent accretion suppression with a fixed value in \sref{sec:bh_sfr}. We show the gas and wind morphology and profiles, as well as the cooling flow properties in \sref{sec:geo_pro}. We summarize our conclusions and discuss the implications for constructing physically motivated BH subgrid prescriptions for galaxy and cosmological simulations in \sref{sec:conclusions}.

\section{Methodology}
\label{S:methods}
We carry out idealized, isolated-galaxy simulations designed to represent M87$^*$- and Sgr A$^*$-like systems with the {\small GIZMO}\footnote{A public version of {\small GIZMO} is available at \href{http://www.tapir.caltech.edu/\~phopkins/Site/GIZMO.html}{\textit{http://www.tapir.caltech.edu/$\sim$phopkins/Site/GIZMO.html}}.} code \citep{Hopkins2015mfm}. The calculations employ the mesh-less finite-mass (MFM) formulation of {\small GIZMO}, a Lagrangian, mesh-free Godunov scheme that combines properties of traditional grid-based approaches with those of smoothed-particle hydrodynamics (SPH). The numerical algorithms and their validation have been presented extensively in previous work, including the treatment of hydrodynamics and self-gravity \citep{Hopkins2015mfm} and magnetohydrodynamics \citep[MHD;][]{Hopkins+2016mhd}. The magnetic-field initialization is described in Section~2.4. We adopt the same initial magnetic configuration used in the multi-zone calculations of \cite{Cho+2024} to facilitate a direct and internally consistent comparison between the two sets of simulations.

\subsection{Stellar Physics}\label{a:sfb}

All simulations adopt the FIRE-2 prescriptions for the interstellar medium (ISM), star formation, and stellar feedback, as described and numerically validated in detail by \citet{Hopkins+FIRE2018}. Gas cooling and heating are followed over the temperature range $10-10^{10}\,{\rm K}$, including photo-electric and photo-ionization heating together with collisional, Compton, fine-structure, recombination, atomic, and molecular cooling processes.

Star formation is implemented using a sink-particle prescription and is restricted to gas that is molecular, self-shielding, locally self-gravitating, and denser than $n>1000\,{\rm cm^{-3}}$ \citep{Hopkins+FIRE2018}. Each newly formed star particle represents a single stellar population and inherits the metallicity of its progenitor gas particle. The associated stellar evolution and feedback quantities, including supernova rates, stellar mass-loss rates, and stellar spectra, are IMF-averaged predictions from {\small STARBURST99} \citep{Leitherer+Starburst99_1999}, assuming the \citet{Kroupa+IMF2002} IMF.

The stellar feedback implementation includes three principal channels. First, radiative feedback accounts for photo-ionization and photo-electric heating as well as radiation pressure from both single and multiple scattering, with the radiation field divided into five bands: ionizing, FUV, NUV, optical-NIR, and IR. Second, winds from OB and AGB stars continuously return mass, metals, momentum, and energy to the surrounding gas. Third, both Type II and Type Ia SNe, including prompt and delayed Type Ia populations, occur according to tabulated event rates and inject the corresponding mass, metals, momentum, and energy into the local ISM. Magnetohydrodynamics (MHD) is included in all simulations.

\subsection{Black hole accretion}

Black hole accretion is modeled based on the Bondi prescription,
\begin{align}\label{eq:bondi}
\dot{M}_{\rm Bondi} = \frac{\alpha \pi \rho G^2 M_{\rm BH}^2}{(c_s^2 + \Delta v^2)^{3/2}},
\end{align}
where $\rho$ is the local gas density, $c_s$ is the mean sound speed, $\Delta v$ is the mean velocity of the gas relative to the BH, and $\alpha$ is a dimensionless prefactor. Unless otherwise noted, we adopt $\alpha=1$, corresponding to adiabatic Bondi accretion with $\gamma=5/3$. These quantities are evaluated by averaging within a kernel enclosing 96 effective neighbors, subject to a minimum radius of 50 comoving pc/$h$.

The actual BH accretion rate is scaled from the Bondi rate by a prefactor $\kappa^{\rm hot}_{\rm cho}$,
following the multi-zone GRMHD-informed model of \citet{Cho+2024,Cho+2025}:
\begin{align}\label{eq:kaappa_cho}
\dot{M}_{\rm BH}&=\kappa^{\rm hot}_{\rm cho} \dot{M}_{\rm Bondi}\notag\\ 
\kappa^{\rm hot}_{\rm cho}&\equiv \left(\frac{r_B}{6r_g}\right)^{-1/2}=\left(\frac{6(c_{s,\infty}^2+\Delta v^2)}{c^2}\right)^{1/2}, 
\end{align}
where $r_B\equiv G M_{\rm BH}/(c_{s,\infty}^2+\Delta v^2)$ is the Bondi radius, and $r_g\equiv G M_{\rm BH}/c^2$ is the gravitational radius. 
We note that for an M87*-like nuclear environment, where the gas temperature is a few times $10^7$ K, the suppression factor would be of order $\kappa^{\rm hot}_{\rm cho}\sim 0.01$. For the environment surrounding Sgr A*, assuming the virial temperature of the Milky Way, this would be of order $\kappa^{\rm hot}_{\rm cho}\sim 0.001$. The measured values of $\kappa^{\rm hot}_{\rm cho}$ are discussed in detail in \sref{sec:kappa_effect}.  

\subsection{AGN feedback}\label{subsec:AGNfeedback}

AGN feedback is treated as an anisotropic wind that predominantly carries kinetic energy using the particle spawning method developed in \citet{Torrey+2020,Su+2021,Weinberger+2023,Su+JetHaloMass2024}, in which new gas resolution elements are created directly around the central BH. The mass assigned to these newly spawned elements is removed from the BH particle, thereby explicitly conserving mass. This procedure allows the properties of the injected outflow to be specified more directly, with substantially less sensitivity to the details of the local neighbor search, while also permitting the launched material to be maintained at higher numerical resolution. The implementation adopted here closely follows that of \citet{Torrey+2020}, who applied a similar prescription to model broad absorption line (BAL) winds in disk galaxies. 

The mass resolution of the spawned gas elements is given in \tref{tab:model}. To prevent premature numerical mixing, these elements are not allowed to de-refine by merging with ordinary gas elements until their velocity has fallen below 10\% of the launch velocity \footnote{A set of simulations was used to test the 10\% threshold for merging spawned particles, and this criterion should be sufficiently stringent with respect to its effects on galaxies and black hole accretion. A more physical and computationally efficient criterion for merging and derefining spawned particles is currently being developed for the \textsc{STARFORGE} project \citep{Grudic2024_starforge} and will be tested and adopted in future work.}. Outflow particles are generated in pairs: once the accumulated injected mass reaches twice the target particle mass, two particles are spawned simultaneously. When spawned, the two particles are launched in exactly opposite directions, guaranteeing exact momentum conservation. Each particle is initially placed at a random position on a sphere of radius $r_0$, where $r_0$ is taken to be the smaller of 10 pc and half the distance from the BH to its nearest gas element. Its launch radial velocity and temperature are assigned according to the following prescriptions.

The mass and energy fluxes of the outflow are given by
\begin{align} 
\dot{M}^{\rm tot}_{\rm wind}=(1-\kappa^{\rm hot}_{\rm cho}) \dot{M}_{\rm Bondi}.
\end{align}
The wind mass flux is divided among angular bins according to:
\begin{align} 
\dot{M}^{\rm}_{\rm wind}(\theta)= f_{\rm mass} (\theta)\dot{M}^{\rm tot}_{\rm wind}.
\end{align}
Note that, as we are using the particle spawning method, the probability of spawning in a specific bin is
\begin{align} 
p(\theta)\propto \frac{f_{\rm mass}(\theta)}{m_{\rm wind}(\theta)} .
\end{align}
The quantity $m_{\rm wind}(\theta)$ specifies the target mass resolution of the spawned particles and can be set as a function of $\theta$. The values of $m_{\rm wind}(\theta)$ adopted in this work are summarized in \tref{tab:model}.

The total feedback-energy flux is given by:


\begin{eqnarray} 
\dot{E}^{\rm tot}_{\rm wind} &~=~&\eta \dot{M}_{\rm BH} c^2\notag\\
\dot{E}_{\rm thermal}(\theta)&~=~&f_{\rm thermal} (\theta)\dot{E}^{\rm tot}_{\rm wind}\notag\\
\dot{E}_{\rm kinetic}(\theta)&~=~&f_{\rm kinetic} (\theta)\dot{E}^{\rm tot}_{\rm wind}.
\end{eqnarray}
All angular information on accretion and feedback ($f_{\rm mass}(\theta)$, $f_{\rm thermal}(\theta)$, $f_{\rm kinetic}(\theta)$) is extracted directly at the Bondi radius from the rapidly spinning ($a_*=0.9$) GRMHD simulations of \citet{Cho+2025} with $R_B\approx 2\times 10^5\,r_g$. As described in \citet{Cho+2025}, at the Bondi radius, the $\theta-$integrated thermal and kinetic energy outflow rates are comparable $\int \dot{E}_{\rm thermal}\,d\theta\approx \int \dot{E}_{\rm kinetic}\,d\theta\approx \dot{E}^{\rm tot}_{\rm wind}/2$. In this work, we assume the overall efficiency $\eta=0.3$ following the feedback efficiency of $a_*=0.9$ of \citet{Cho+2025}.

The specific energy is therefore:
\begin{align}\label{eq:spec}
\epsilon_{\rm thermal/kinetic}(\theta)&\equiv\frac{\dot{E}_{\rm thermal/kinetic}(\theta)}{\dot{M}_{\rm wind}(\theta)}\notag\\
&=\frac{\eta\kappa^{\rm hot}_{\rm cho} f_{\rm thermal/kinetic}(\theta)c^2}{(1-\kappa^{\rm hot}_{\rm cho})f_{\rm mass} (\theta)}
\end{align}

For each outflow rate, the feedback particle properties can be derived as follows. For the kinetic component,
\begin{align} 
\dot{E}_{\rm kinetic}(\theta)=\frac{1}{2} \dot{M}_{\rm wind}(\theta) V_{\rm wind}(\theta) ^2,
\end{align}
so
\begin{align} 
V_{\rm wind}(\theta) &= \left(\frac{2\eta f_{\rm kinetic} (\theta) \dot{M}_{\rm BH}}{f_{\rm mass} (\theta)\dot{M}^{\rm tot}_{\rm wind}}\right)^{1/2} c\notag\\
&= \left(\frac{2\eta \kappa^{\rm hot}_{\rm cho}f_{\rm kinetic} (\theta)}{(1-\kappa^{\rm hot}_{\rm cho}) f_{\rm mass} (\theta) }\right)^{1/2} c.
\end{align}

Meanwhile, the temperature is
\begin{align} 
T(\theta)=\frac{\mu m_p}{k_B} (\gamma-1)\epsilon_{\rm thermal}(\theta).
\end{align}

\begin{figure}
    \centering
 \includegraphics[width=7.5cm]{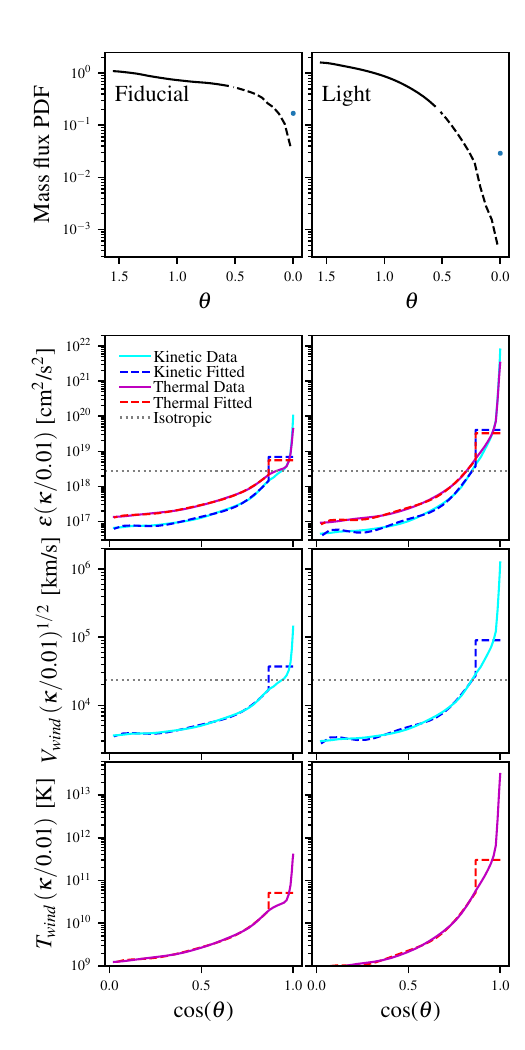}    

    \caption{Angular PDF of the wind mass flux (first row) and angular profiles of the specific energy ($\epsilon$; second row), wind velocity ($V_{\rm wind}$; third row), and temperature ($T_{\rm wind}$; fourth row). The mass/energy flux PDF for $\theta>30^\circ$ (=0.52~rad) is extracted directly from \cite{Cho+2025}. For $\theta<30^\circ$, all wind is launched along the pole, with the corresponding probability represented by the point. For the quantities in rows 2--4, our implementation uses fitted profiles, with $\theta<30^\circ$ treated as a single bin and assigned one averaged value. The specific kinetic energy and velocity of the wind in the Isotropic runs are represented by the dashed black line.}
    \label{fig:ftheta}
\end{figure}

The models tested in this work are summarized below and consist of one isotropic run and two variations of the flux distributions, as shown in the two columns of \fref{fig:ftheta}. The first row shows the mass flux PDF. For $\theta>30^\circ$, the PDF is extracted directly from \cite{Cho+2025}. For $\theta<30^\circ$, wind is launched along the pole, with the corresponding probability represented by the point. The second row shows the angular distribution of specific energy given by \Eqref{eq:spec}. The dashed lines show the profiles implemented in the code, while the solid lines show the original distribution obtained from \cite{Cho+2025}. Again, the deviations happen within $30^\circ$. The third and fourth rows show the corresponding angular dependence of the wind velocity and temperature, respectively. 
 
We note that for $\theta>30^\circ$, we closely follow the distribution obtained by \cite{Cho+2025}. For $\theta<30^\circ$, we average this distribution over the angular range, treating it as a single bin. This effectively reduces the specific energy near the pole while preserving the total energy and mass fluxes, thereby avoiding extremely high velocities that would be challenging to handle in MHD simulations. We detail the variations among the runs below.
 
\begin{itemize}
\item {\bf Isotropic:\space} As a reference, we first consider an isotropic kinetic wind model similar to that of \cite{Su+2025Bridging}, with $f_{\rm mass}(\theta)$ and $f_{\rm kinetic}(\theta)$ both set to constant values chosen to preserve the total mass and energy fluxes of the distributions from \cite{Cho+2025}. We assign this energy flux entirely to kinetic energy and set the wind temperature uniformly to $10^7$ K, following \cite{Su+2025Bridging}. The associated thermal energy flux is much smaller than the kinetic energy flux. The feedback is therefore characterized by angle-independent efficiencies. The specific kinetic energy and velocity of the wind are represented by the dashed black line in \fref{fig:ftheta}.

We emphasize that this reference model changes more than just the wind geometry. Relative to the GRMHD-informed models, isotropizing the feedback replaces the angular distributions of mass and energy loading with a single angularly averaged value for each quantity, thereby also removing the angular variation in specific energy. We therefore use the Isotropic model to test the consequences of replacing the full GRMHD-informed angular feedback structure with an isotropic, angularly averaged prescription.

\item {\bf VTh-Fiducial:\space} The fiducial case, shown in the left column, averages the distributions within the polar region ($\theta<30^\circ$) into a single angular bin to avoid excessively high velocities while preserving the total mass and energy fluxes. In addition, jets originating from this region are constrained to be launched within $1^\circ$ of the pole.

\item {\bf VTh-Light:\space} The third variation, shown in the right column, follows the implementation of the second variation but artificially decreases $f_{\rm mass}(\theta)$ around the pole. Specifically, while keeping the same $f_{\rm thermal/kinetic}(\theta)$, we use the inflow distribution, $f_{\rm mass}(\theta)=f^{\rm in}_{\rm mass}(\theta)$, instead of the outflow distribution, $f_{\rm mass}(\theta)=f^{\rm output}_{\rm mass}(\theta)$, obtained from \cite{Cho+2025} as outflows. Because the inflow is minimal near the pole, this effectively reduces the mass loading in the polar region while keeping the energy flux fixed. The resulting specific energy in the polar region is therefore much higher.

\end{itemize}

\begin{figure*}
    \centering
    \includegraphics[width=17cm]{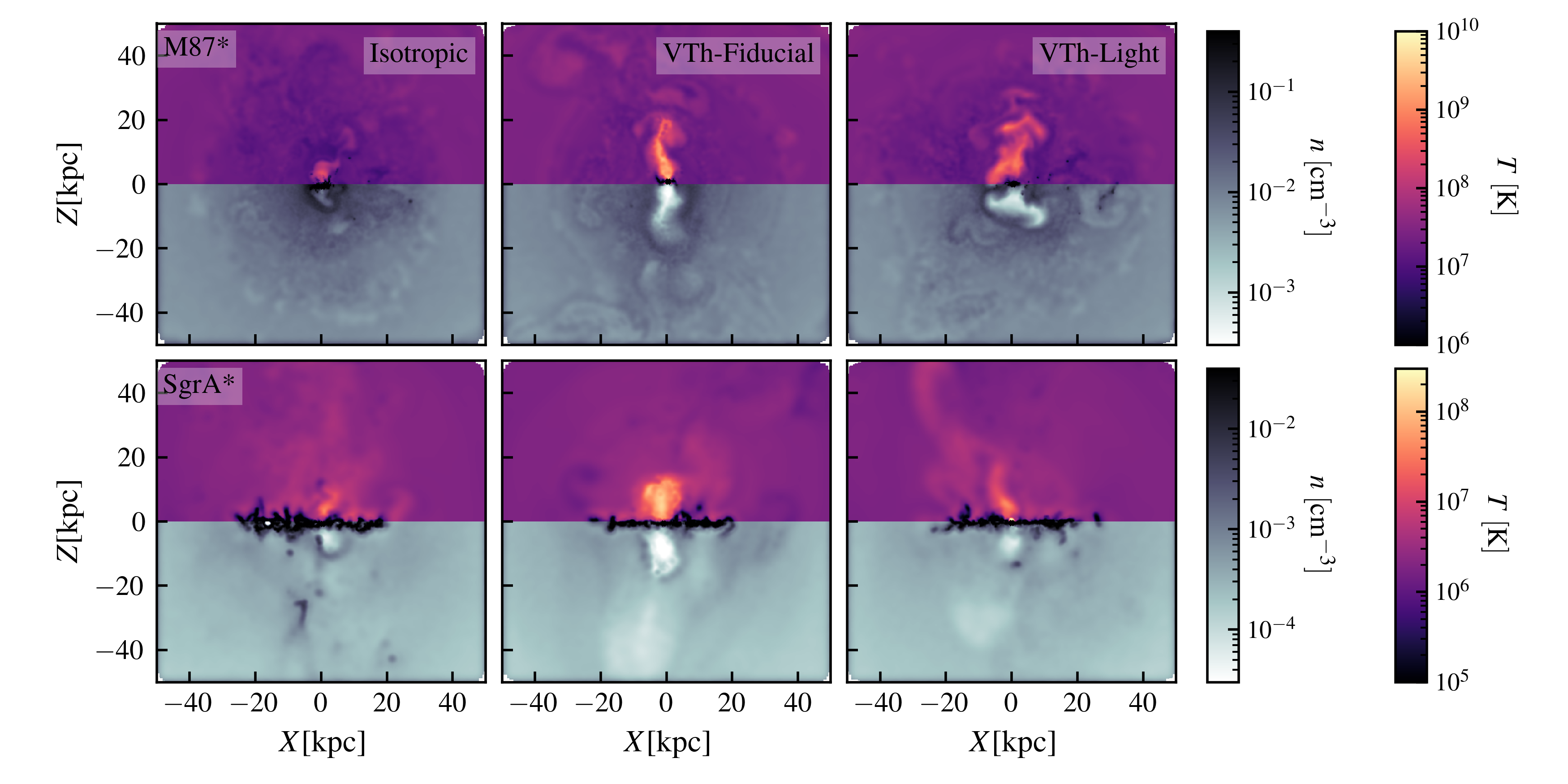}

    \caption{Gas temperature (upper half) and density (lower half) distributions at peak feedback activity after 1 Gyr for the M87* analogue and after 2 Gyr for the Sgr A* analogue. The three columns show the isotropic wind (Isotropic, left), fiducial jet (VTh-Fiducial, center), and light jet (VTh-Light, right), respectively. The isotropic wind produces a less extended heated region and remains mostly within 10 kpc, while runs with an angular distribution of flux produce jets that reach larger radii. The cocoon of the light jet broadens at smaller radii, whereas the fiducial jet is more collimated and remains coherent out to larger radii.} 
    \label{fig:wind}
\end{figure*}

For these three models, \fref{fig:wind} illustrates the CGM gas morphology through density and temperature slices of the simulations. The snapshots are selected at a local peak of a wind episode after 1 Gyr in the M87* case and after 2 Gyr in the Sgr A* case. As noted above, the isotropic wind remains mostly within 10 kpc, while the GRMHD-informed angular feedback models produce polar outflows that reach substantially larger radii. In particular, the jet cocoon in the run with the light jet broadens at smaller radii, whereas the fiducial jet is more collimated and remains coherent out to larger radii.

The spawned wind material is initialized with a magnetic field of $10^{-5}\,{\rm G}$ oriented along the $z$-direction. Under these conditions, the magnetic contributions to the injected energy flux remain subdominant to the kinetic component. This choice is motivated by the multi-zone calculations of \citet{Cho+2024}, in which the outflow at the Bondi radius is primarily carried by fluid energy rather than magnetic energy. We therefore adopt a weak magnetic field for the AGN outflow. Fields below $10^{-4}\,{\rm G}$ are dynamically unimportant in this context and have a negligible influence on the resulting evolution \citep{Su+2021,Su+JetHaloMass2024}. We nevertheless retain a finite, non-zero field in the injected material because the ambient galaxy is magnetized, and setting the wind field identically to zero would introduce an artificial region of anomalously weak magnetization.

The list of runs is detailed in Table~2.

\begin{table*}
\begin{center}
\caption{The initial conditions used in this work correspond to those in \cite{Su+2025Bridging}}
\label{tab:ic}
 \resizebox{18.cm}{!}
 {
 \movetableright=-2.2in
 \begin{tabular}{l|C| c c |c|c c c| c |c c| c c| c c| c c c }
 \hline
\hline
&&\multicolumn{2}{c|}{\underline{Resolution}}&\underline{BH}&\multicolumn{3}{c|}{\underline{DM halo}}&&\multicolumn{2}{c|}{\underline{Stellar Bulge}}&\multicolumn{2}{c|}{\underline{Stellar Disk}}&\multicolumn{2}{c|}{\underline{Gas Disk}}&\multicolumn{3}{c}{\underline{Gas Halo}}  \\
$\,\,\,\,$Model &R_{200} &$\epsilon_g$ &$m_g$        &$M_{\rm BH}$&$M_{\rm halo}$   &$r_{dh}$            &$\rho_{0}$    &$M_{\rm bar}$    &$M_b$ 
                 &$a$          &$M_d$        & $r_d$             &$M_{gd}$       &$r_{gd}$         &$M_{gh}$         &$r_{gh}$ &$\beta$    \\
                &(kpc) &(pc)        &(M$_\odot$) &(M$_\odot$)  &(M$_\odot$)      & (kpc)           &(g cm$^{-3}$)           &(M$_\odot$)      &(M$_\odot$) 
                  &(kpc)        &(M$_\odot$)  &(kpc)            &(M$_\odot$)    &(kpc)            &(M$_\odot$)      &(kpc)&\\
\hline
$\,\,\,\,$M87*        &1306&1       &2e4   &6.5e9        &2.0e14           &448             &2.0e-26              &5.2e13           &6.9e11   
                     &3.97       &-          &-                &-            &-                &5.1e13           &0.93   &0.33     \\  
$\,\,\,\,$Sgr A* -light CGM  &241&1       & 8e3 &4.3e6          &1.5e12           &20             &5.8e-25              &1.2e11           &1.5e10   
                     &1.0       &5.0e10          &3.0                &5.0e9            &6.0                &4.6e10           &20  &0.5      \\

\hline 
\end{tabular}

}

\end{center}
\begin{flushleft}
(1) Model name. 
(2) $R_{200}$: Radius within which the mean enclosed density is 200 times the critical density of the Universe.
(3) $\epsilon_g$: Minimum gravitational force softening for gas (the softening for gas in all simulations is adaptive, and matched to the hydrodynamic resolution; here, we quote the minimum Plummer equivalent softening).
(4) $m_g$: Gas mass (resolution element). There is a resolution gradient for M87* analog runs, 
so its $m_g$ is the mass of the highest resolution elements.
(5)$M_{\rm BH}$: BH mass.
(6) $M_{\rm halo}$: Dark matter halo mass within $R_{\rm vir}$. 
(7) $r_{dh}$: NFW halo scale radius, which yields a central concentration parameter
$c \sim$ 12 [MW halo] \& 3 [for the M87 halo].
(8) $\rho_{\rm 0}$: The density normalization for the NFW profile.
(9) $M_{\rm bar}$: Total baryonic mass within $R_{\rm vir}$. 
(10) $M_b$: Bulge mass.
(11) $a$: Bulge Hernquist-profile scale-length.
(12) $M_d$ : Stellar disk mass.
(13) $r_d$ : Stellar disk exponential scale-length.
(14) $M_{gd}$: Gas disk mass. 
(15) $r_{gd}$: Gas disk exponential scale-length.
(16) $M_{gh}$: Hydrostatic gas halo mass within $R_{\rm vir}$. 
(17) $r_{gh}$: Hydrostatic gas halo $\beta$ profile scale-length.
(18) $\beta$: Hydrostatic gas halo $\beta$ profile $\beta$.
\end{flushleft}
\end{table*}

\begin{table*}
\begin{center}
\caption{AGN Feedback Models and Simulation Variants}
 \label{tab:model}
\resizebox{18.cm}{!}
 {
  \movetableright=-1.56in
\begin{tabular}{lc|c|cccccc|cccc}

\hline
&& & \multicolumn{6}{c|}{Feedback variations} & \multicolumn{4}{c}{Results } \\
\hline
\multicolumn{2}{c|}{Model} & $\Delta T$  &  $m_{\rm wind}^{<30^o}$ &  $m_{\rm wind}^{30^o-45^o}$ & $m_{\rm wind}^{45^o-60^o}$ & $m_{\rm wind}^{>60^o}$ & Ceiling    & Altered Weight& $M^{\rm 1.8/2.5 Gyr}_{\rm BH}$ &$M^{\rm 1.8/2.5 Gyr}_{\rm *}$ & $\langle\dot{M}_{\rm BH}\rangle$ & $\langle\dot{M}_{\rm *}\rangle$ \\
&&(Gyr) & (${\rm M_\odot}$) & (${\rm M_\odot}$) & (${\rm M_\odot}$)& (${\rm M_\odot}$) & $<30^o$  & $<30^o$ & (${\rm M_\odot}$) & (${\rm M_\odot}$)& (${\rm M_\odot}$ yr$^{-1}$) & (${\rm M_\odot}$ yr$^{-1}$) \\
\hline
\multicolumn{2}{l|}{\underline{\textcolor{blue}{M87*}}}&&&&&&&&\\
Iso&M87*-Isotropic         &1.9& 6e3 &6e3 &6e3 &6e3 &N/A  &N/A &6.50e9&7.05e11 &1.3e-3&9.2\\
Fiducial&M87*-VTh-Fiducial        &1.9& 6e2 &2e3 &6e3 &6e3 &capped  & no&6.50e9&6.91e11&1.2e-3&0.70\\
Light&M87*-VTh-Light  &1.9& 6e2 &2e3 &6e3 &6e3 &capped  & light &6.50e9&6.91e11&1.4e-3&0.58\\

\hline
\multicolumn{2}{l|}{\underline{\textcolor{blue}{Sgr A* -light CGM}}}&&&&&&&&&\\
Iso&SgrA*-Isotropic              &3.5& 1e3 &1e3 &1e3 &1e3 &N/A  &N/A  &4.30e6&6.72e10&3.1e-6&0.83\\
Fiducial&SgrA*-VTh-Fiducial       &3.5& 1e2 &3e2 &1e3 &1e3 &capped  &no &4.31e6&6.67e10 &3.4e-6&0.66\\
Light&SgrA*-VTh-Light       &3.5& 1e2 &3e2 &1e3 &1e3 &capped  &light &4.30e6&6.67e10 &4.1e-6&0.62\\
\hline 
\hline
\end{tabular}
}
\end{center}
\begin{flushleft}
 
(1)Model name. 
(2) $\Delta T$: Duration of the runs. 
(3,4,5,6) $m_{\rm wind}$: Mass resolution of the spawned AGN wind particles in each $\theta$ range. 
(7) Whether the jet velocity and temperature are capped for $\theta<30^\circ$.
(8) Whether and how the angular distribution of the flux is modified.
(9,10) $M^{\rm 1.8/2.5\,Gyr}_{\rm BH/*}$: BH and stellar masses after 1.8 Gyr for the M87$^*$ analog runs and after 2.5 Gyr for the Sgr A$^*$ analog runs.
(11,12) $\langle\dot{M}_{\rm BH}\rangle$  $\langle\dot{M}_{\rm *}\rangle$: The time-averaged BH accretion and star formation rates from 0.5-1.5 Gyr for M87* analog runs and from 0.5-2.5 Gyr for Sgr A* analog runs.

\end{flushleft}
\end{table*}

\subsection{Initial conditions}\label{a:ic}

We construct galaxy-scale initial conditions representative of an M87$^*$-like massive elliptical system \citep{2008MNRAS.388.1062C,2012MNRAS.421..635F,2016MNRAS.457..421O} and a Milky Way analog hosting an Sgr A$^*$-like BH \citep{Miller+MW2013,Miller+MW2015,Gupta+2017}. For the Milky Way model, we adopt the circumgalactic medium (CGM) normalizations corresponding to the `light CGM' configuration in \cite{Su+2025Bridging}. This model uses the CGM profile inferred by \citet{Miller+MW2015} and is denoted the Sgr A$^*$-light CGM model. 

The dark matter (DM), stellar, BH, and gaseous components are generated following the general equilibrium procedures described by \citet{1999MNRAS.307..162S,2000MNRAS.312..859S} and adopted in our previous simulations \citep{Su+2019,Su+2020,Su+2021,Su+JetHaloMass2024}. The DM component is initialized as a spherical and isotropic NFW halo \citep{1996ApJ...462..563N}. For the Sgr A$^*$ and M87$^*$ analogs, respectively, the halo masses enclosed within $R_{200}$ are $1.5\times10^{12}$ and $1.97\times10^{14}\,{\rm M_\odot}$, with corresponding NFW scale radii of 20 and 448 kpc. The stellar bulges follow \citet{1990ApJ...356..359H} profiles, with masses of $1.5\times10^{10}$ and $6.87\times10^{11}\,{\rm M_\odot}$ and scale radii of 1 and 3.97 kpc for the Sgr A$^*$ and M87$^*$ systems, respectively. The central BH masses are $M_{\rm BH}=4.3\times10^6\,{\rm M_\odot}$ for Sgr A$^*$ and $6.5\times10^9\,{\rm M_\odot}$ for M87$^*$.

The gaseous halos are initialized in hydrostatic equilibrium with $\beta$-model density profiles. The gas masses enclosed within $R_{200}$, defined as the radius within which the mean total mass density is 200 times the critical density of the Universe, are $4.6\times10^{10}$ and $4.7\times10^{13}\,{\rm M_\odot}$ for the fiducial Sgr A$^*$-light CGM and M87$^*$ models, respectively. The corresponding $\beta$ parameters are $(0.5,0.33)$ and the core radii are $(20,0.93)$ kpc. As the simulations evolve, these initially hydrostatic atmospheres relax toward rotating cooling-flow configurations \citep{Stern+coolingflow2024}. In the M87$^*$ analog, the halo gas is assigned an initial angular frequency of $0.1\Omega_K$, where $\Omega_K$ denotes the Keplerian angular frequency. The gas, therefore, remains predominantly supported by thermal pressure, as expected for a halo of this mass. The Sgr A$^*$ analog is initialized with a somewhat larger rotation rate, $0.3\Omega_K$. Beyond 30 kpc, however, its azimuthal velocity declines as $v_\phi\propto r^{-1}$, producing an approximately constant-specific-angular-momentum profile. The Milky Way analog additionally contains exponential, rotationally supported stellar and gaseous disks. Their radial scale lengths are 3 and 6 kpc, respectively, while both components have a vertical scale height of 0.3 kpc.

The initial metallicity distribution is prescribed as
\begin{equation}
Z(r)=Z_\odot\left[Z_{\rm out}
+\frac{1-Z_{\rm out}}
{1+(r/R_c)^{\beta_z}}\right],
\end{equation}
with $(Z_{\rm out},R_c,\beta_z)=(0.05,20\,{\rm kpc},1.5)$ for the Sgr A$^*$ analog and $(0.26,64\,{\rm kpc},2)$ for the M87$^*$ analog. In this work, we do not impose an artificial temperature floor around M87$^*$ or Sgr A$^*$, as we did in \cite{Su+2025Bridging}.

The initial magnetic configuration changes from predominantly poloidal in the central region to predominantly toroidal at larger radii. Within $\sim1$ kpc, the field is initialized with $\beta\sim1$, while beyond 10 kpc, the toroidal component approaches $\beta=1000$. Here, the poloidal component follows the functional form adopted by \citet{Cho+2024}. For the M87$^*$ analog, we use exactly the same initial magnetic-field configuration as \citet{Cho+2024}, allowing the galaxy-scale calculation to be compared as directly as possible with their multi-zone simulations. The Sgr A$^*$ model adopts the same radial and geometric prescription, but rescales the overall field strength to account for its different thermodynamic conditions while retaining approximately the same maximum $\beta$ in the central kiloparsec.

The fiducial gas mass resolution in the Sgr A$^*$ simulations is $8\times10^3\,{\rm M_\odot}$ throughout the region within 244 kpc. For the M87$^*$ analog, the highest gas mass resolution is $2\times10^4\,{\rm M_\odot}$. We employ a hierarchical super-Lagrangian refinement prescription \citep[e.g.,][and our previous series of papers]{AnglesAlcazar+Zoom_2021,Su+2023,Su+JetHaloMass2024} to concentrate resolution toward the halo center. The gas mass resolution is finest within $r_{\rm 3d}<10\,{\rm kpc}$. For $r_{\rm 3d}>10\,{\rm kpc}$, the target gas-element mass increases approximately linearly with radius as $m_g=1.7\times10^4[1+(r_{\rm 3d}-10\,{\rm kpc})/(6\,{\rm kpc})]\,{\rm M_\odot}$, reaching $\sim4\times10^6\,{\rm M_\odot}$ at $R_{200}$.

\section{Black hole accretion and star formation rate}\label{sec:bh_sfr}

\subsection{The effect of angular feedback structure}
\label{sec:geo}
\begin{figure*}
    \centering
    \includegraphics[width=17cm]{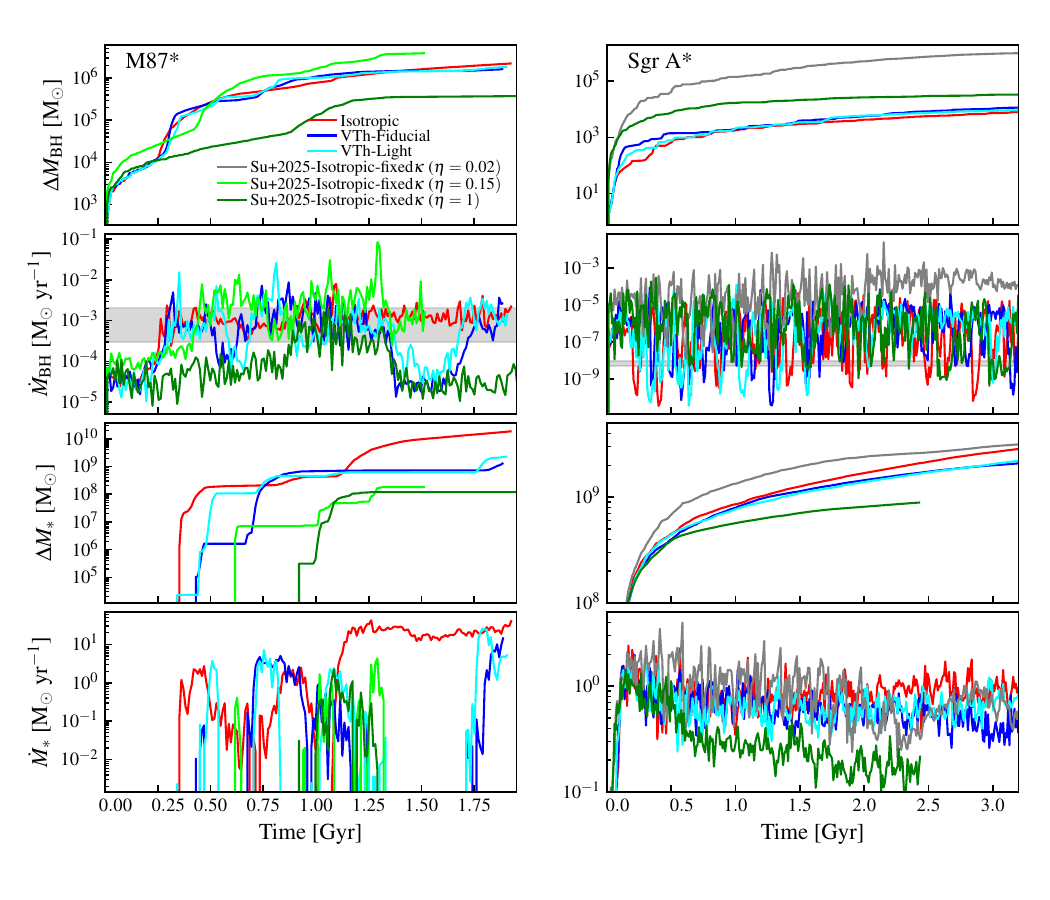}

    \caption{The net increase in black hole mass (first row), the black hole accretion rate (second row), the net stellar mass formed since the start of the simulation (third row), and the star formation rate (fourth row).
   Shown are results for the three primary models in this work: Isotropic (isotropic wind, red), VTh-Fiducial (fiducial jet, blue), and VTh-Light (lighter jet, cyan). For comparison, runs from \cite{Su+2025Bridging} with different feedback efficiencies are shown in gray, lime, and green.
     The gray shaded region highlights the black hole accretion rate range inferred from EHT observations \citep{EHTm87_8_2021,EHTMW_5_2022}. In both the M87* and Sgr A* cases, the isotropic wind is less efficient at quenching star formation, resulting in higher stellar masses. The effect of angular feedback structure on BH accretion is comparatively small. In the Sgr A* case, all the new runs in this work have lower black hole accretion rates, which are more consistent with the accretion-rate range inferred from EHT observations. }
    \label{fig:bh_sfr}
\end{figure*}

\begin{figure*}
    \centering
    \includegraphics[width=17cm]{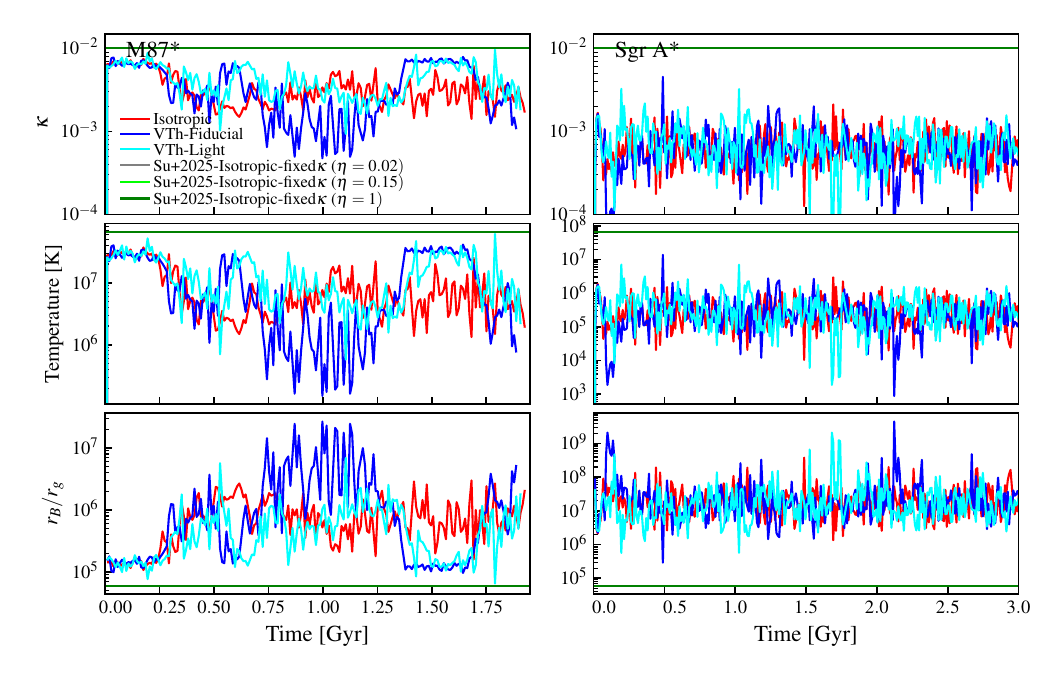}

    \caption{Time evolution of the accretion suppression factor (top, $\kappa^{\rm hot}_{\rm cho} \equiv \dot{M}{\rm BH}/\dot{M}{\rm Bondi}$), the corresponding effective gas temperature (middle), and the Bondi radius (bottom). Shown are results for the three primary models in this work: Isotropic (isotropic wind, red), VTh-Fiducial (fiducial jet, blue), and VTh-Light (lighter jet, cyan). For comparison, runs from \cite{Su+2025Bridging} with different feedback efficiencies are shown in gray, lime, and green. In the Sgr A* case, the effective temperature is around $10^5$ K, resulting in a suppression factor orders of magnitude below $0.01$. In the M87* case, the suppression factor is closer to $0.01$. }
    \label{fig:kappa}
\end{figure*}

\fref{fig:bh_sfr} shows the black hole accretion rate, the net increase in black hole mass, the star formation rate, and the net stellar mass formed since the start of the simulation. The gray shaded region highlights the black hole accretion rate range inferred from EHT observations \citep{EHTm87_8_2021,EHTMW_5_2022}.

As emphasized in Section~\ref{subsec:AGNfeedback}, the Isotropic comparison does not vary geometric collimation alone. Relative to the GRMHD-informed models, it also
removes the correlated angular variations in mass loading, kinetic and thermal-energy fluxes, and specific energy. We therefore interpret the differences below as the effect of the full \emph{angular feedback structure}, rather than of geometry alone.

Comparing the `Isotropic', `VTh-Fiducial', and `VTh-Light' runs for both the M87* and Sgr A* cases reveals a few trends. With the same feedback efficiency in all three variants, isotropic injection is the least efficient at quenching star formation. Compared with the runs that retain the GRMHD-informed angular feedback structure, the isotropic run forms roughly ten times as much stellar mass in the M87$^*$ case and roughly 1.5 times as much in the Sgr A$^*$ case. 

This is because the GRMHD-informed angular feedback structure naturally produces higher specific energy near the poles and lower specific energy near the midplane, as shown in \fref{fig:ftheta}, effectively yielding a jet plus wind configuration. In contrast, isotropic injection produces uniformly moderate specific energy. Injection with high specific energy is more efficient at suppressing cooling flows and quenching star formation \citep{Su+2021}.

However, once the GRMHD-informed angular feedback structure is 
included, the detailed form of the specific energy profile in the `jet core', where the jet velocity is $\gtrsim 3\times10^{4}\,{\rm km\,s^{-1}}$, has a smaller effect on the resulting star formation. As detailed in \cite{Su+2021}, winds with high specific energy enhance quenching in two ways. First, material with high specific energy has a much longer cooling time and, therefore, loses less energy through radiation before affecting the surrounding gas. Second, at the same energy flux, a jet with lower specific energy has a higher momentum flux, remains narrow, and reaches larger radii, whereas a jet with higher specific energy expands and becomes quasi-isotropic at smaller radii. A jet with a velocity of $\sim 3\times10^{4}\,{\rm km\,s^{-1}}$ is already sufficiently effective in both respects, so making it even lighter does not significantly change the result.

Changing the detailed polar structure of the jet and wind has a smaller
effect on BH accretion than replacing the full GRMHD-informed angular
feedback structure with an isotropic prescription. All variants show very similar net increases in black hole mass. Interestingly, for M87*, the runs with a feedback efficiency of $\eta=0.3$, corresponding to a black hole spin of $a_*=0.9$, have accretion rates within the range suggested by the EHT (the gray shaded region). The only difference is that the run with isotropic energy injection maintains a steadier black hole accretion rate throughout the simulation. This is again because an isotropic wind, despite its higher mass loading, has uniformly lower specific energy and spreads over a much larger solid angle than a jet. It therefore does not propagate as far and regulates accretion closer to the black hole, on scales with shorter dynamical times. 

For Sgr A*, the black hole accretion rates are also broadly consistent with the values suggested by the EHT \citep[$\lesssim10^{-8} M_\odot {\rm yr}^{-1}$,][]{EHTMW_5_2022}. Interestingly, the agreement is better than for any of the runs with different feedback efficiencies in \cite{Su+2025Bridging}. We will compare the runs in \cite{Su+2025Bridging} with those in this work briefly in \sref{sec:kappa_effect} and in detail in \sref{a:comparison}. 

When comparing our simulated star formation rates with observations, we find that all our models lie within the range of published Milky Way estimates, $\sim0.67$--$1.9\,M_\odot\,{\rm yr}^{-1}$ \citep{Chomiuk+2011MWSFR,Quintana+2025MWSFR}, for most of the simulation time. For M87, UV observations give an SFR upper limit of $0.1\,M_\odot\,{\rm yr}^{-1}$ \citep{Hoffer+2012M87SFR}.

\subsection{The effect of nuclear-environment-dependent black hole accretion suppression}\label{sec:kappa_effect}

To quantify the effect of using a suppression factor $\kappa^{\rm hot}_{\rm cho}$ that depends on the local environment in the BH vicinity, including the gas temperature and velocity, we also include runs from our previous paper \citep{Su+2025Bridging}, which used a constant accretion suppression factor of $\kappa = 0.01$.

We highlight that, in addition to the difference in feedback efficiency, the runs in our previous paper \citep{Su+2025Bridging} differ from those in this work majorly in the following ways. First, the suppression relative to Bondi accretion was fixed at $\kappa=0.01$, rather than determined by a relation that depends on the local environment in the BH vicinity (\Eqref{eq:kaappa_cho}). Second, feedback was injected fully isotropically and entirely as kinetic energy. Finally, the previous M87* runs imposed a temperature floor of $\sim10^6$ K within roughly 1 kpc. Among these differences, the change in black hole accretion suppression is the most important.

Despite these differences in modeling, a few comparisons can be made between the previous and current runs. All new runs in this paper use a feedback efficiency of $\eta=0.3$. The new M87* runs accrete at rates roughly between those of the previous runs with $\eta=0.15$ and $\eta=1$, consistent with the idea that higher black hole feedback efficiency leads to lower accretion rates. For Sgr A*, however, the new runs accrete at lower rates than even the previous run with the highest feedback efficiency ($\eta=1$). This brings the black hole accretion rates into much better agreement with the values suggested by the EHT.

The main reason lies in the suppression factor for black hole accretion, $\kappa^{\rm hot}_{\rm cho}$. \fref{fig:kappa} shows the time evolution of this suppression factor, together with the corresponding effective temperature and effective $r_{\rm Bondi}/r_g$.  We define the effective temperature as $T_{\rm eff} = m(c_s^2 + \Delta v^2)/\gamma k$, with $m$ denoting the mean particle mass. We see that $\kappa^{\rm hot}_{\rm cho}$ is much lower than 0.01 at the ambient temperatures in the Sgr A* runs. Thus, despite the lower feedback efficiency, the new runs have lower accretion rates because accretion in the Sgr A* environment is more strongly suppressed than we previously estimated. As shown in \fref{fig:kappa}, 
this discrepancy introduced by adopting a uniform suppression factor of
$\kappa=0.01$ is less pronounced in the M87* case because $\kappa^{\rm hot}_{\rm cho}$ in the much hotter nuclear environment of the more massive halo is closer to 0.01. This further confirms that $\kappa^{\rm hot}_{\rm cho}$ depends strongly on halo mass. A comparison of the stellar mass formed in the new and previous runs also reveals an interesting trend.
In the Sgr A* case, the stellar mass formed in the new runs with a feedback efficiency of $\eta=0.3$  falls between the values for the previous runs with $\eta=0.02$ and $\eta=1$, consistent with the picture that higher feedback efficiency leads to more efficient quenching. However, this simple picture does not hold for M87*, where all new runs form more stars than the previous runs, even when the new runs have higher feedback efficiency. This difference is primarily due to the removal of the temperature floor imposed within 1 kpc in the previous runs. Because many stars form within 1 kpc, this floor suppressed the star formation rate in the previous runs relative to the new runs.

We note that, for the M87$^*$ runs in this work, the fiducial model exhibits much larger variations in the suppression factor $\kappa^{\rm hot}_{\rm cho}$ and the corresponding local effective temperature than the light jet model, with episodic excursions toward lower values. This is consistent with the fact that the fiducial jet, with its lower specific energy, heats the vicinity of the black hole less efficiently. However, these differences do not translate into changes in net black hole growth or star formation.

\section{Gas inflows and profiles}\label{sec:geo_pro}

\begin{figure*}
    \centering
    \includegraphics[width=17cm]{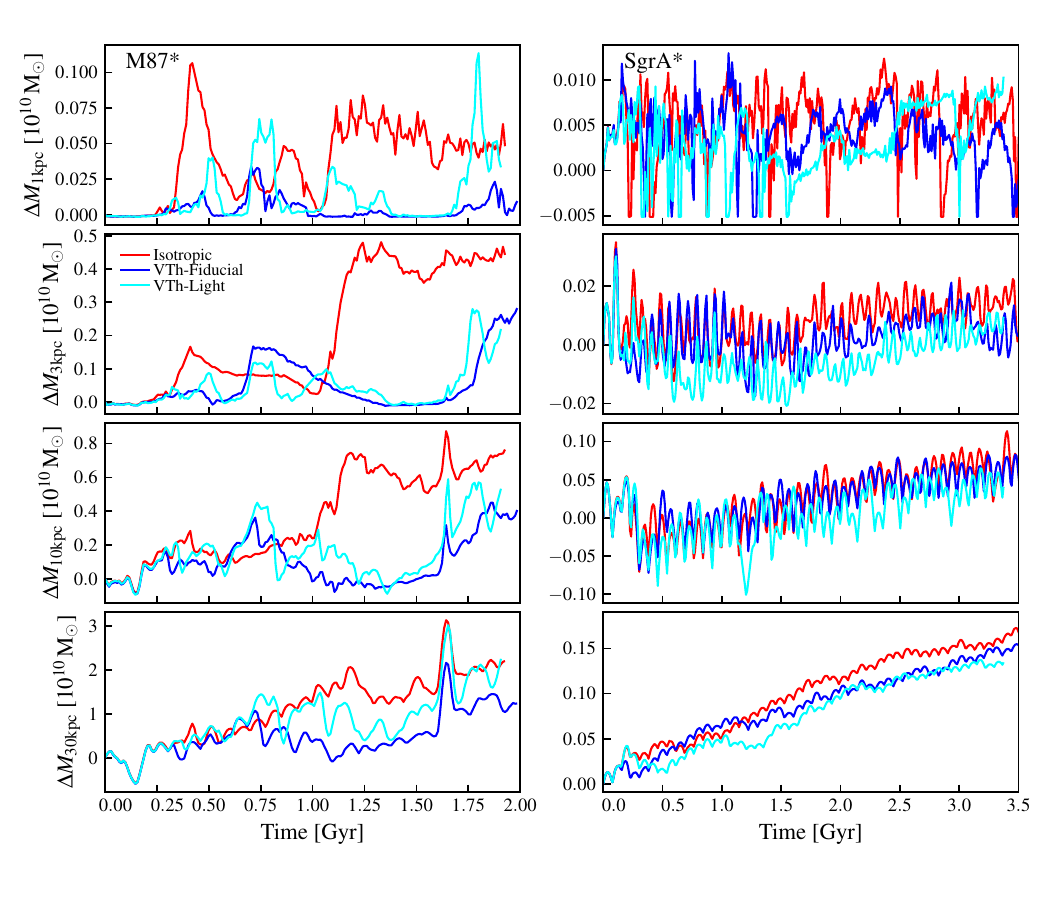}

    \caption{Change in the gas mass enclosed within 1, 3, 10, and 30 kpc relative to the initial values. Shown are results for the three primary models in this work: Isotropic (isotropic wind, red), VTh-Fiducial (fiducial jet, blue), and VTh-Light (lighter jet, cyan). At the same feedback efficiency, the isotropic wind suppresses the cooling flow less effectively, resulting in a higher gas mass on average.}
    \label{fig:inflow}
\end{figure*}
\begin{figure*}
    \centering
    \includegraphics[width=17cm]{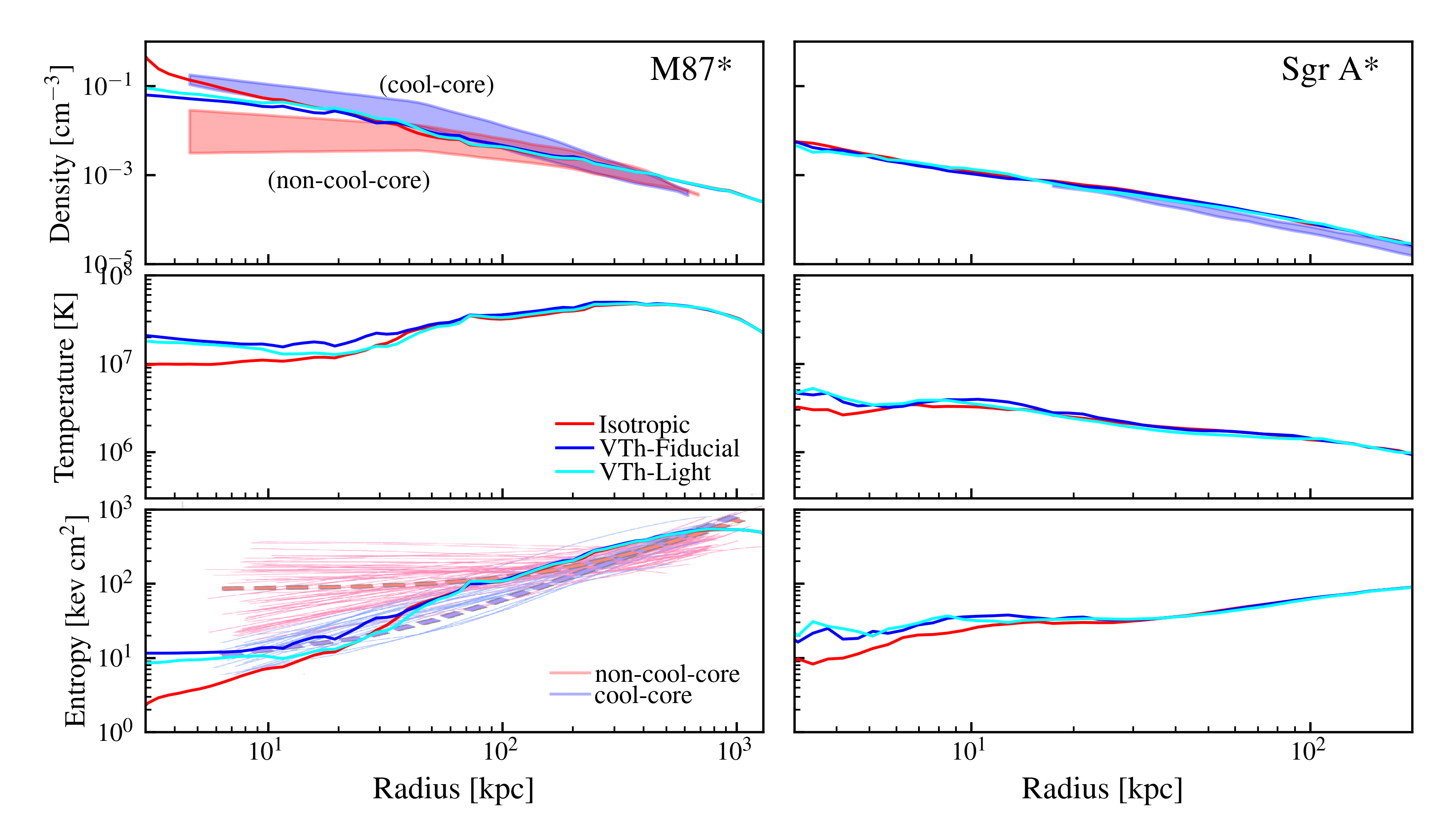}

    \caption{Radial distributions of gas density, temperature, and entropy in the M87* and Sgr A* analog simulations. For M87$^*$, profiles are averaged over 1.45--1.5 Gyr using the 0.5--7 keV X-ray luminosity as the weight. For Sgr A$^*$, profiles are averaged over the same interval using the mass of gas at $T > 10^5$ K as the weight, since the gas is generally too cool to emit strongly in this X-ray band.  Shown are results for the three primary models in this work: Isotropic (isotropic wind, red), VTh-Fiducial (fiducial jet, blue), and VTh-Light (lighter jet, cyan). Observational reference profiles are shown by the shaded bands. For M87*, these correspond to cool-core clusters in blue and non-cool-core clusters in red \citep{McDonald+cluster2013}, rescaled to account for differences in halo mass. For Sgr A*, the shading indicates Milky Way profiles \citep{Miller+MW2015}. In both systems, isotropic winds produce denser, cooler gas with lower entropy in the central region.}
    \label{fig:etd}
\end{figure*}

\begin{figure*}
    \centering
    \includegraphics[width=17cm]{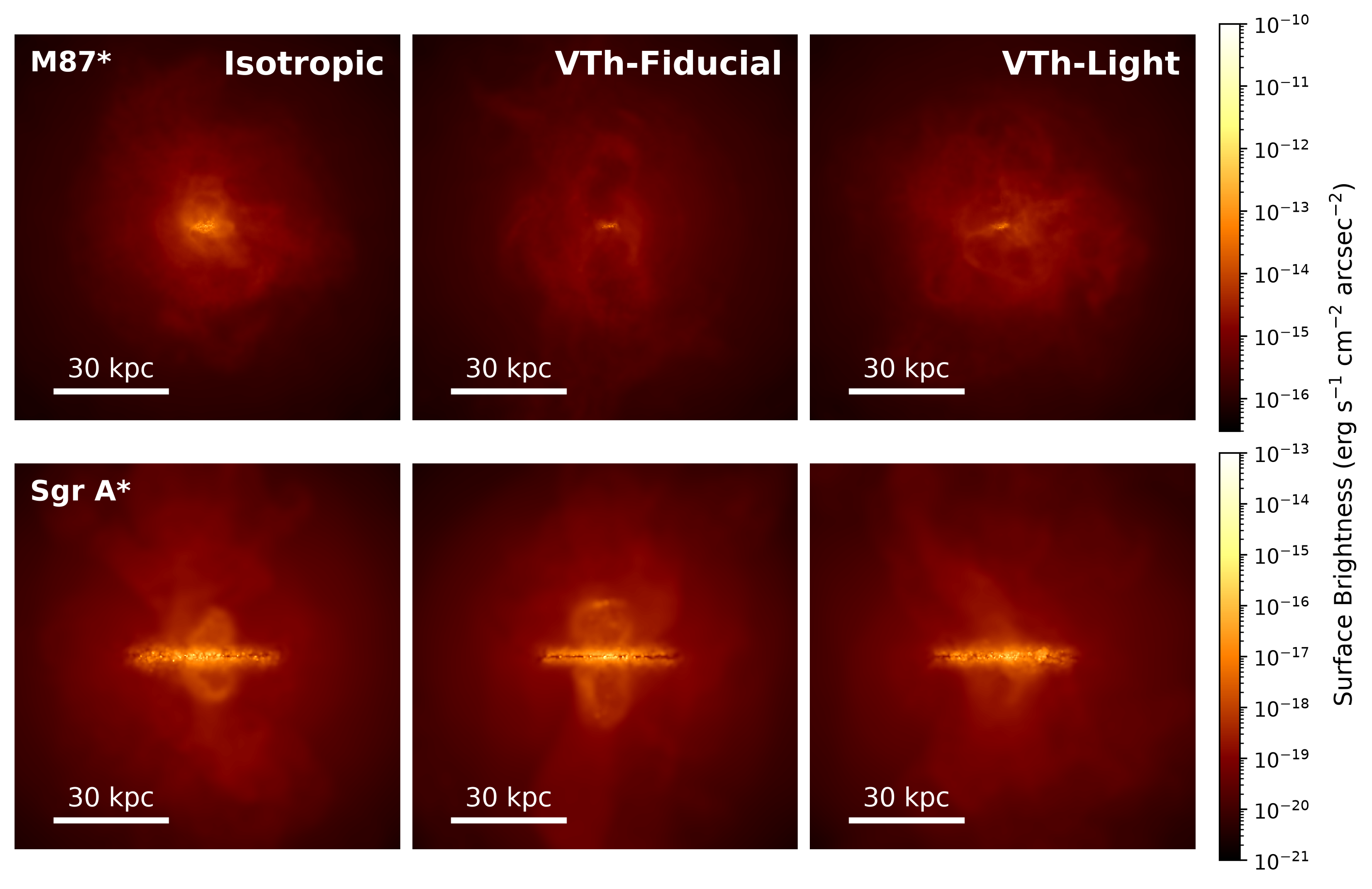}

    \caption{X-ray surface brightness in the 0.5--7 keV band for all runs at the same snapshots as in \fref{fig:wind}, when the accretion rates reach local maxima. The three columns show the isotropic wind (Isotropic, left), fiducial jet (VTh-Fiducial, center), and light jet (VTh-Light, right), respectively. The X-ray bubbles are most prominent in VTh-Fiducial runs. The surface brightness of the Sgr A* runs broadly matches the observed range for the eROSITA bubbles, and the Sgr A* VTh-Fiducial run also resembles the observed bubbles in shape and size at the snapshot shown \citep{erosita_bubble_2020}. }
    \label{fig:x_ray}
\end{figure*}

In section \sref{sec:geo}, we attribute the more effective quenching in the GRMHD-informed angular feedback models relative to the isotropic model to their ability to launch high-specific-energy outflows that suppress cooling flows more efficiently. Fig. 5 shows changes in the gas mass enclosed within radii of 1, 3, 10, and 30 kpc relative to the initial values. In all cases, the enclosed gas mass at each radius is higher on average in the isotropic wind run during most of the simulation, consistent with a stronger cooling flow and less efficient quenching. A consistent trend can be seen in \fref{fig:etd}, which shows the X-ray luminosity-weighted density, temperature, and entropy profiles for the M87* analogue runs and the mass-weighted profiles restricted to gas hotter than $10^5$ K for the Sgr A* runs. The isotropic wind case generally produces higher densities, lower temperatures, and lower entropies within $\sim30$ kpc, comparable to the radius typically reached by the jets, as shown in \fref{fig:wind}. These profiles are therefore slightly more characteristic of cool-core systems, consistent with the stronger cooling flows. Once the angular dependence is included, the fiducial and light jet cases produce similar density, temperature, and entropy profiles.

In \fref{fig:x_ray}, we also show the X-ray surface brightness in the 0.5--7 keV band for all runs at the same snapshots as in \fref{fig:wind}, when the accretion rates reach local maxima. The X-ray bubbles are most prominent in the runs with our fiducial jet model (VTh-Fiducial), because the heavier jets produce more collimated cocoons. The surface brightness of the Sgr A* runs broadly matches the observed range for the eROSITA bubbles, and the Sgr A* VTh-Fiducial run also resembles the observed bubbles in shape and size at the snapshot shown \citep{erosita_bubble_2020}.

\section{Summary and Conclusions}
\label{sec:conclusions}

In this work, we have developed and tested a galaxy-scale implementation of black-hole accretion and feedback designed to connect directly to the multi-zone GRMHD calculations of \cite{Cho+2024,Cho+2025}. We apply this framework to idealized isolated-galaxy simulations representative of both an M87$^\ast$-like massive elliptical and a Milky Way/Sgr A$^\ast$-like system. The accretion model modifies the classical Bondi rate using the GRMHD-motivated suppression  factor $\kappa^{\rm hot}_{\rm cho}$, while the associated feedback is launched as an anisotropic wind whose feedback power, along with angular distributions of mass, kinetic energy, and thermal energy, is informed by the same GRMHD nuclear-scale calculations. In this way, a consistent underlying prescription is used across systems spanning several orders of magnitude in black-hole and halo mass, without separately tuning an effective accretion or feedback efficiency for each host.

Our results indicate that the angular structure of the nuclear outflow can have an important effect on its coupling to the surrounding galaxy. In the M87$^\ast$-like system, the spatially uniform and GRMHD-informed angular
feedback models produce substantially different star-formation histories, even when their long-term BH accretion rates are comparatively similar. In particular, the GRMHD-motivated anisotropic models more effectively suppress the late-time buildup of stellar mass than the isotropic reference model. This demonstrates that the impact of AGN feedback cannot, in general, be characterized only by its total injected energy or momentum. Retaining the correlated angular variations in mass flux, energy flux, and specific energy can strongly affect where the outflow thermalizes and how efficiently it couples to cooling gas.
 
 At the same time, the galaxy-scale evolution appears to be less sensitive to the uncertain details of the most highly collimated polar component than it is to the distinction between a spatially uniform prescription and the full GRMHD-informed angular feedback structure.

Our fiducial GRMHD-informed jet (VTh-Fiducial) and lighter jet (VTh-Light) models differ substantially in their assumed polar mass loading. In the latter, the polar mass loading is reduced while the energy loading is held fixed, producing a much larger specific energy in the polar outflow. Nevertheless, the global histories of black-hole growth and star formation in these two models remain much more similar to one another than either is to the isotropic case. However, the GRMHD-informed jets are `heavier' in the polar region and therefore produce more collimated cocoons. This also leads to more prominent X-ray cavities than in the other model variants, and in the Sgr A* case, a closer resemblance to the eROSITA bubbles \citep{erosita_bubble_2020}. This suggests that, for a given energy outflow rate, reproducing the broad angular structure of the GRMHD outflow may be as important as (or even more important than) specifying the detailed mass loading of the unresolved jet core for galaxy-scale modeling.

A second important result is that the fraction of the Bondi inflow reaching the black hole is not well described by a universal constant. In our implementation,
\begin{equation}
    \dot{M}_{\rm BH} = \kappa^{\rm hot}_{\rm cho}\dot{M}_{\rm Bondi},
\end{equation}
with $\kappa^{\rm hot}_{\rm cho}$ determined by the ratio of the Bondi and gravitational radii, and therefore by the thermodynamic and kinematic state of the gas surrounding the black hole. The resulting suppression factor varies both with time and between the M87$^\ast$ and Sgr A$^\ast$ analogs. The M87$^\ast$-like simulations typically occupy a different range of $\kappa^{\rm hot}_{\rm cho}$, gas temperature, and $r_{\rm B}/r_g$ than the Sgr A$^\ast$-like simulations, demonstrating that the effective small-scale accretion efficiency responds to the resolved local environment in the BH vicinity rather than being an externally imposed parameter.

This nuclear-environment dependence can have a large effect on predicted black-hole growth. In the Sgr A$^\ast$-like calculations, simulations using the self-consistent $\kappa^{\rm hot}_{\rm cho}$ produce substantially different cumulative black-hole growth from otherwise similar calculations in which the suppression factor is fixed to $\kappa=0.01$. The stronger suppression of accretion yields a black hole accretion rate that agrees better with the value inferred by the EHT \citep{EHTm87_8_2021,EHTMW_5_2022}. By contrast, the corresponding differences in the global star-formation histories are more modest. Thus, a subgrid accretion prescription that yields an approximately reasonable galaxy-scale star-formation rate does not necessarily simultaneously produce the correct black-hole growth history. This also illustrates that black-hole self-regulation and regulation of the host galaxy are related but distinct constraints on feedback models.

Because the suppression factor depends on the local jet angle and jet power, it is not a single universal constant but a quantity that can vary with the state of the inflowing gas. This matters for implementation in cosmological hydrodynamical simulations, which typically adopt one fixed suppression factor for all black holes. If the suppression factor instead varied from galaxy to galaxy according to local jet and gas conditions, black holes in galaxies of the same stellar mass could follow different growth histories, which would increase the scatter in black hole mass at fixed stellar mass. \citet{Porras-ValverdeForbes+2026} show that the faint-end slope of the quenched stellar mass function depends jointly on the scatter in black hole mass, $\sigma_{\rm BH}$, and on how steeply the quenching rate scales with black hole mass. At large $\sigma_{\rm BH}$, quenching prescriptions with very different scalings produce similar faint-end slopes, so simulations with too little scatter must adopt a less aggressive quenching prescription to match observations. Nuclear-environment-dependent accretion suppression is therefore a physically motivated way to raise $\sigma_{\rm BH}$ in cosmological simulations without retuning the feedback prescription. Testing this idea would require a cosmological calculation that evolves the suppression factor self-consistently for a population of black holes.

Taken together, these results support a genuinely multiscale picture of black-hole feedback. Physics on small scales leaves an imprint on galactic scales through the angular structure and specific energy of the outflow, while the thermodynamic state of the host galaxy feeds back onto nuclear scales by determining the effective fraction of Bondi inflow that reaches the black hole. The same GRMHD-informed prescription, therefore, produces different accretion and feedback regimes in the M87$^\ast$ and Sgr A$^\ast$ analogs without requiring separate phenomenological efficiencies for the two systems. More broadly, our calculations suggest that subgrid models based on a fixed Bondi suppression factor and isotropic energy injection may miss two distinct pieces of the coupling between black holes and their hosts: the nuclear-environment dependence of accretion and the angular feedback structure of the resulting outflow.

Several extensions will be important in future work. The simulations presented here are idealized isolated systems, and a broader exploration of halo mass, gas supply, angular momentum, magnetic-field structure, and cosmological inflow will be required to determine how generally these conclusions apply. In particular, the Milky Way calculations were constructed to test whether black-hole feedback can redistribute or remove enough circumgalactic gas to affect the inferred missing-baryon budget, but the present results do not yet establish a quantitative conclusion on that question. Future calculations that follow the radial redistribution and thermodynamic evolution of the CGM in greater detail should provide a direct test of this possibility. Nevertheless, the present results demonstrate that physically motivated information from horizon-to-Bondi-scale calculations can be incorporated into galaxy-scale simulations in a way that changes both black-hole growth and the macroscopic response of the host, providing a path toward more predictive models of AGN feedback across scales.

We also note that the GRMHD-informed accretion-feedback model used in this work is derived from a set of multi-zone GRMHD simulations limited to low accretion rates, in which radiative cooling is neglected. We nevertheless assume that this model applies universally throughout our simulations, regardless of the temperature and density around the black hole. In reality, a separate prescription for cold or super-Eddington accretion may be necessary, and we leave its development to future work.

\section*{Acknowledgements}

KYS and CAFG were supported by NSF through grants AST-2108230, AST-2307327 and AST-2606015; by NASA through grants 80NSSC22k0809, 80NSSC22K1124 and 80NSSC24K1224; by STScI through grant JWST-AR-03252.001-A; and by BSF through grant \#2024262.  PN, HC, RN, and CP were partially supported by the Black Hole Initiative at Harvard University, which is funded by the Gordon and Betty Moore Foundation grant 8273, and the John Templeton Foundation grant 61497. The opinions expressed in this publication are those of the authors and do not necessarily reflect the views of the Moore or Templeton Foundation. 
  AJPV gratefully acknowledges the support of the National Science Foundation Astronomy \& Astrophysics Postdoctoral Fellowship Award Number 2502826.
  The simulations are done on Frontera with allocation AST22010, and Bridges-2 with Access allocations TG-PHY220027 \& TGPHY220047. Support for PFH was provided by NSF Research Grants 1911233, 20009234, 2108318, NSF CA-REER grant 1455342, NASA grants 80NSSC18K0562, HST-AR-15800.
 This research also used resources provided to CP by the Los Alamos National Laboratory Institutional Computing Program, which is supported by the U.S. Department of Energy National Nuclear Security Administration under Contract No. 89233218CNA000001. We acknowledge the use of ChatGPT-6 during the final stages of manuscript preparation to identify typos and grammatical errors, improve sentence clarity, and assist with literature searches.  We thank FIRE and LtU collaboration for useful discussions.  

\section*{Data Availability Statement}
A public release of the \textsc{GIZMO} code is available at
\url{http://www.tapir.caltech.edu/~phopkins/Site/GIZMO.html}.
The complete simulation outputs are not publicly archived because of their large storage requirements. The processed datasets underlying the figures are available from the corresponding author upon reasonable request for scientific purposes.

\appendix
\section{Comparison with our previous work}\label{a:comparison}
\begin{table*}
\begin{center}
\caption{AGN Feedback Models and Simulation Variants}
 \label{tab:ic2}
\resizebox{18.cm}{!}
 {
  \movetableright=-1.2in
\begin{tabular}{lc|c|ccc|cccc}

\hline
&& & \multicolumn{3}{c|}{Feedback variations} & \multicolumn{4}{c}{Results } \\
\hline
\multicolumn{2}{c|}{Model} & $\Delta T$  &  $m_{\rm wind}$ & $\alpha$ &  $\kappa$ & $M^{\rm 2.4 Gyr}_{\rm BH}$ &$M^{\rm 2.4 Gyr}_{\rm *}$ & $\langle\dot{M}_{\rm BH}\rangle$ & $\langle\dot{M}_{\rm *}\rangle$ \\
&&(Gyr) & (${\rm M_\odot}$) & Bondi & &  (${\rm M_\odot}$) & (${\rm M_\odot}$)& (${\rm M_\odot}$ yr$^{-1}$) & (${\rm M_\odot}$ yr$^{-1}$) \\
\hline
\multicolumn{2}{l|}{\underline{\textcolor{blue}{Sgr A* -light CGM}}}&&&&&\\
ISO      &SgrA*-Isotropic                    &2.5& 1e3 &1 &$\kappa^{\rm hot}_{\rm cho}$ &4.30e6&6.71e10 &3.2e-6&0.83\\
ISO2     &SgrA*-Iso-fixed$\kappa$            &2.5& 1e3 &1 &0.01 &4.31e6&6.65e10&5.7e-6&0.54\\
ISO3     &SgrA*-Iso-fixed$\kappa$-lowres     &2.5& 8e3 &1 &0.01 &4.33e6&6.60e10&1.3e-5&0.35\\
ISO4     &SgrA*-Iso-fixed$\kappa$-lowres-4x  &2.5& 8e3 &4 &0.01&4.33e6&6.60e10&1.4e-5 &0.35\\
\hline 
\hline
\end{tabular}
}
\end{center}
\begin{flushleft}
 
(1)Model name. 
(2) $\Delta T$: Duration of the runs. 
(3) $m_{\rm wind}$: Mass resolution of the spawned AGN wind particles. 
(4) $\alpha$, the overall prefactor for the Bondi accretion rate.
(5) Whether $\kappa^{\rm hot}_{\rm cho}$ depends on the local environment in the BH vicinity or is fixed at $0.01$.

(6,7) $M^{\rm 2.4\,Gyr}_{\rm BH/*}$: BH and stellar masses after 2.4 Gyr.
(8,9) $\langle\dot{M}_{\rm BH}\rangle$  $\langle\dot{M}_{\rm *}\rangle$: The averaged BH accretion rate from 0.5-2.4 Gyr.
\end{flushleft}
\end{table*}
\begin{figure}
    \centering
    \includegraphics[width=17cm]{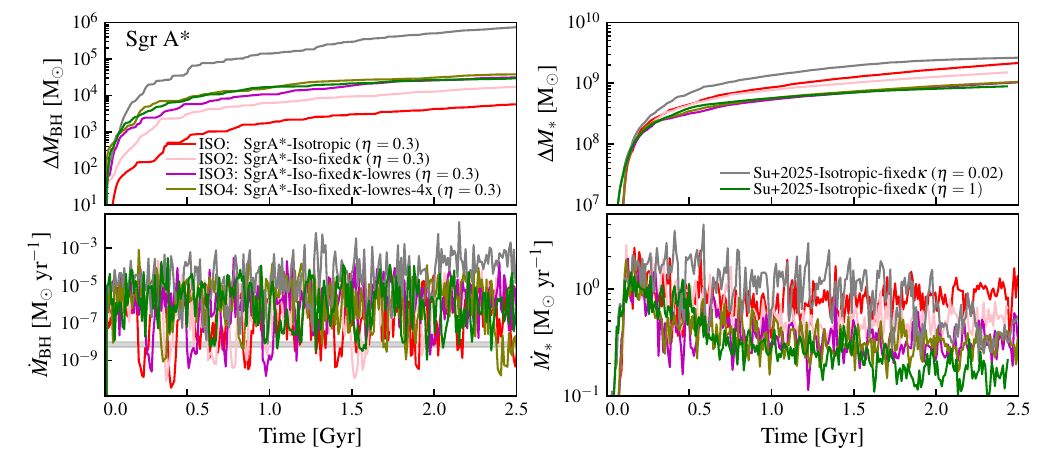}

    \caption{The black hole accretion rate, net increase in black hole mass, star formation rate, and net stellar mass formed since the start of the simulation for the four runs listed in \tref{tab:ic2}. ISO4 follows the implementation of \cite{Su+2025Bridging} and yields a black hole accretion rate between those of the previous runs with higher and lower feedback efficiencies. Allowing $\kappa^{\rm hot}_{\rm cho}$ to vary with the local environment in the BH vicinity lowers the black hole accretion rate. Increasing the mass resolution of the wind particles also slightly lowers the accretion rate.}
    \label{fig:res}
\end{figure}

In \sref{sec:kappa_effect}, we noted the major differences in the implementation of black hole accretion and feedback between this work and \cite{Su+2025Bridging}. For the Sgr A* runs, the extensive list of differences is as follows:

First, the previous runs fixed the suppression relative to Bondi accretion at $\kappa=0.01$, rather than using a relation that depends on the local environment in the BH vicinity (\Eqref{eq:kaappa_cho}). Second, feedback was injected fully isotropically and entirely as kinetic energy. Third, the previous runs had lower resolution. Finally, \cite{Su+2025Bridging} adopted $\alpha=4$ in \Eqref{eq:bondi}, whereas this work uses $\alpha=1$.

In \sref{sec:kappa_effect}, we attributed the generally lower black hole accretion rates in the new Sgr A* runs to the nuclear-environment dependence of $\kappa^{\rm hot}_{\rm cho}$, which is much lower than 0.01 for Sgr A*. Here, we test this interpretation quantitatively through a sequence of runs that progressively restore the earlier prescriptions. The runs are listed in \tref{tab:ic2}. We consider four variants: ISO is the run analyzed in the main part of the paper; ISO2 fixes the accretion suppression factor at $\kappa=0.01$; ISO3 additionally lowers the resolution to that of \cite{Su+2025Bridging}; and ISO4 further restores the $\alpha=4$ prefactor in the Bondi accretion rate.

The resulting black hole accretion rate, net increase in black hole mass, star formation rate, and net stellar mass formed since the start of the simulation are shown in \fref{fig:res}. With the earlier accretion and resolution settings restored in ISO4, the black hole accretion rate lies between those of the \cite{Su+2025Bridging} runs with feedback efficiencies of $\eta=0.02$ and $\eta=1$. Since all four ISO variants use $\eta=0.3$, this result is consistent with the expectation that higher feedback efficiency leads to lower black hole accretion rates.

The difference between ISO and ISO4 is driven primarily by the nuclear-environment dependence of $\kappa^{\rm hot}_{\rm cho}$, as shown by the comparison between ISO and ISO2.

Resolution also plays a role. Comparing ISO2 and ISO3 shows that increasing the mass resolution of the spawned wind particles reduces the black hole accretion rate, although the effect is smaller than that of allowing $\kappa^{\rm hot}_{\rm cho}$ to depend on the local environment in the BH vicinity.

\bibliography{bridging_scales}{}
\bibliographystyle{aasjournal}


\end{CJK}
\end{document}